\documentclass[aoas]{imsart}

\RequirePackage{amsthm,amsmath,amsfonts,amssymb}
\RequirePackage[authoryear]{natbib}
\usepackage[utf8]{inputenc}
\usepackage{amsmath,amssymb}
\usepackage{algorithm}
\usepackage{algorithmic}
\usepackage{xcolor}
\usepackage{graphicx}
\usepackage{hyperref}
\usepackage{appendix}

\startlocaldefs

\endlocaldefs

\begin{document}

\begin{frontmatter}
\title{Expected Points Above Average: A Novel NBA Player Metric Based on Bayesian Hierarchical Modeling}
%\title{A sample article title with some additional note\thanksref{T1}}
\runtitle{Expected Points Above Average}
%\thankstext{T1}{A sample of additional note to the title.}
%%%%%%%%%%%%%%%%%%%%%%%%%%%%%%%%%%%%%%%%%%%%%%%
%% Only one address is permitted per author. %%
%% Only division, organization and e-mail is %%
%% included in the address.                  %%
%% Additional information can be included in %%
\begin{aug}
\author{\fnms{Benjamin} \snm{Williams}\thanksref{m1}},
\author{\fnms{Erin M.} \snm{Schliep}\thanksref{m2}},
\author{\fnms{Bailey K.} \snm{Fosdick}\thanksref{m3}},
\and 
\author{\fnms{Ryan} \snm{Elmore}\thanksref{m1}\ead[label=e1]{Ryan.Elmore@du.edu}}
% \and

\address{\thanksmark{m1}Daniels College of Business\\
Department of Business Information\\ and Analytics\\
University of Denver\\
\printead{e1}}
\address{\thanksmark{m2}Department of Statistics\\
North Carolina State University}
\address{\thanksmark{m3}Department of Biostatistics and Informatics\\Colorado School of Public Health\\University of Colorado, Anschutz Medical Campus}
\end{aug}

\begin{abstract}
%Team and player evaluation in professional sport is extremely important given the financial implications of success/failure. It is especially critical to identify and retain elite shooters in the National Basketball Association (NBA), one of the premier basketball leagues worldwide because the ultimate goal of the game is to score more points than one's opponent. 
In this paper, we propose two novel basketball metrics: ``expected points'' for team-based comparisons and ``expected points above average (EPAA)'' as a player-evaluation tool. Established within the Bayesian hierarchical model framework, teams and players are clustered based on their shooting propensities and abilities using posterior predictive distributions. We illustrate the concepts for the top 100 shot takers over the last decade and offer our metric as an additional metric for evaluating players. We compare our metrics to two traditional NBA player evaluation metrics: player efficiency rating and box plus/minus. Finally, we develop a Shiny web application that allows interested readers to make additional team and player comparisons.  
%both metrics leverage posterior samples from a Bayesian hierarchical modeling framework to cluster teams and players based on their shooting propensities and abilities. 
\end{abstract}


\end{frontmatter}
%%%%%%%%%%%%%%%%%%%%%%%%%%%%%%%%%%%%%%%%%%%%%%
%% Please use \tableofcontents for articles %%
%% with 50 pages and more                   %%
%%%%%%%%%%%%%%%%%%%%%%%%%%%%%%%%%%%%%%%%%%%%%%
%\tableofcontents

%%%%%%%%%%%%%%%%%%%%%%%%%%%%%%%%%%%%%%%%%%%%%%
%%%% Main text entry area:
\section{Introduction}
\label{sec:intro}

Sports analytics research has three primary goals: evaluating individual impact, improving winning, and quantifying components of the game. These pillars often go hand-in-hand; analytical tools quantifying player impact or features of a game can also support changes in game strategy, which may lead to more winning. With these goals in mind, we revisit the classic problem of measuring player contributions for professional basketball players in the National Basketball Association (NBA). Naively, because the objective of basketball is to score more points than the opponent, it may seem that the most valuable players are those who score the most points. However, this approach ignores other aspects of the game, for example the skill(s) of the other players on the team. A better way to evaluate a player might be by their shooting percentage, defined as the number of made shots divided by the total number of shots taken. This approach also falls short since it succumbs to the small sample fallacy, whereby players with few shots may have extremely variable make percentages.

There are several sophisticated metrics by which to evaluate players; two popular metrics are Player Efficiency Rating (PER) and Box Plus/Minus (BPM). PER incorporates a variety of basketball statistics for the offensive and defensive contributions of each player \citep{hollinger}, such as rebounds, assists, and steals. One benefit of PER is that it is measured on a per-minute basis, removing any worry about small sample sizes of playing time for a player. However, a criticism of PER is that it is not interpretable. 

Alternatively, BPM is a regression-based approach and is defined as the points contributed by a player per 100 possessions above an average player \citep{BPM}. BPM incorporates statistics similar to PER--including both offensive and defensive--and considers the position of the player. Although BPM is more interpretable than PER since its units is in points, a major drawback is that it relies on the position of the player, which is defined more ambiguously in the modern NBA \citep{mcmahan}.

\citet{DeshpandeJensen} argue that metrics such as PER and BPM which use statistics such as points, rebounds, assists, etc. fail to account for game context.  With this motivation,  \citet{DeshpandeJensen} propose quantifying the contribution of a player in terms of their team's odds of winning a basketball game, attempting to control for elements of the game such as ``high-leverage" and `low-leverage" situations (when the teams are close in score or far away in score, respectively). Alternatively, \citet{daly} investigated an individual player's impact on the game by modeling how a defensive player effects the shot trajectory of an offensive player, ignoring the location of the field goal itself.

With the goals of quantifying aspects of the game and determining the best strategies for winning, other basketball research has implicitly considered the problem of measuring player performance. \citet{sand_etal_20} quantify the efficiency of different groupings of NBA teammates using a Bayesian spatial modeling approach relying on a metric measuring a player's shot efficiency given which teammates were also on the court. In an attempt to validate known relationships between winning and player performance, and to potentially identify new relationships, \citet{durso22} estimated a multivariate Bayesian network to study how different player variables affect winning.

Previous efforts have also looked at shot location data. \citet{sand_born} used Markov decision processes to analyze shooting strategies by investigating NBA player and shot location data. 
\citet{reich2006spatial} used hierarchical spatial models for shot-chart data for individual players, with one stated goal being to develop a strategy for guarding specific players.

Turning to clustering players based on similar shooting characteristics, \citet{jiao_21} used a Bayesian marked spatial point process model to analyze NBA shot charts and clustered players into groups according to their individual shot taking data.
\citet{hu2021bayesian} utilized a mixture of finite mixtures model to examine the heterogeneity of shot locations of players to find players with similar shooting profiles.  
\citet{yinhushen} extended this work by examining clusters of players via Bayesian inference, which allows for estimating the optimal number of clusters while also estimating cluster membership. Clustering was also the goal of \citet{MunizFlamand} who employed k-Means clustering of player statistics to identify outlying players and player archetypes. Others have gone beyond simply using player statistics to measure player performance, for example \citet{zhang2018clustering} used both player height and weight measurements in addition to playing statistics for constructing player clusters. %They employed a two-stage cluster analysis, along with a discriminant analysis for identifying the clusters and offer the results as new ways of classifying players.

In this paper, we creatively translate a metric for player comparison originally developed for baseball into the basketball arena. We call it \textit{expected points above average} (EPAA). This metric helps teams better identify players that drive winning. Our derivation of the metric also quantifies player and team shooting profiles in a way which offers insight for comparing teams and players. Our work emulates the \textit{runs above average} statistic commonly used in Major League Baseball \citep{winston2022mathletics, isar} to evaluate players based on their expected point contribution to the average team above what an average player would contribute. A parallel metric for evaluating players is called \textit{wins above replacement}, where the statistic captures the increase in the number of wins a player contributes to their team as opposed to if they were replaced with a \textit{replacement player}. \textit{Wins above replacement} has been used in baseball, see \cite{openwar} where the authors used \textit{runs above average to calculate} \textit{wins above replacement}. It also has found use in American football, e.g. \cite{nflwar}, and hockey \citep{hockey-vent}. These metrics are nicely interpretable and have widespread use in professional sports. This led us to adopt the same form for our metric which is a measure of the contribution of a player above the \textit{ average player} as opposed to that of a \textit{replacement player}.  

\begin{figure}
\includegraphics[width=5.5in]{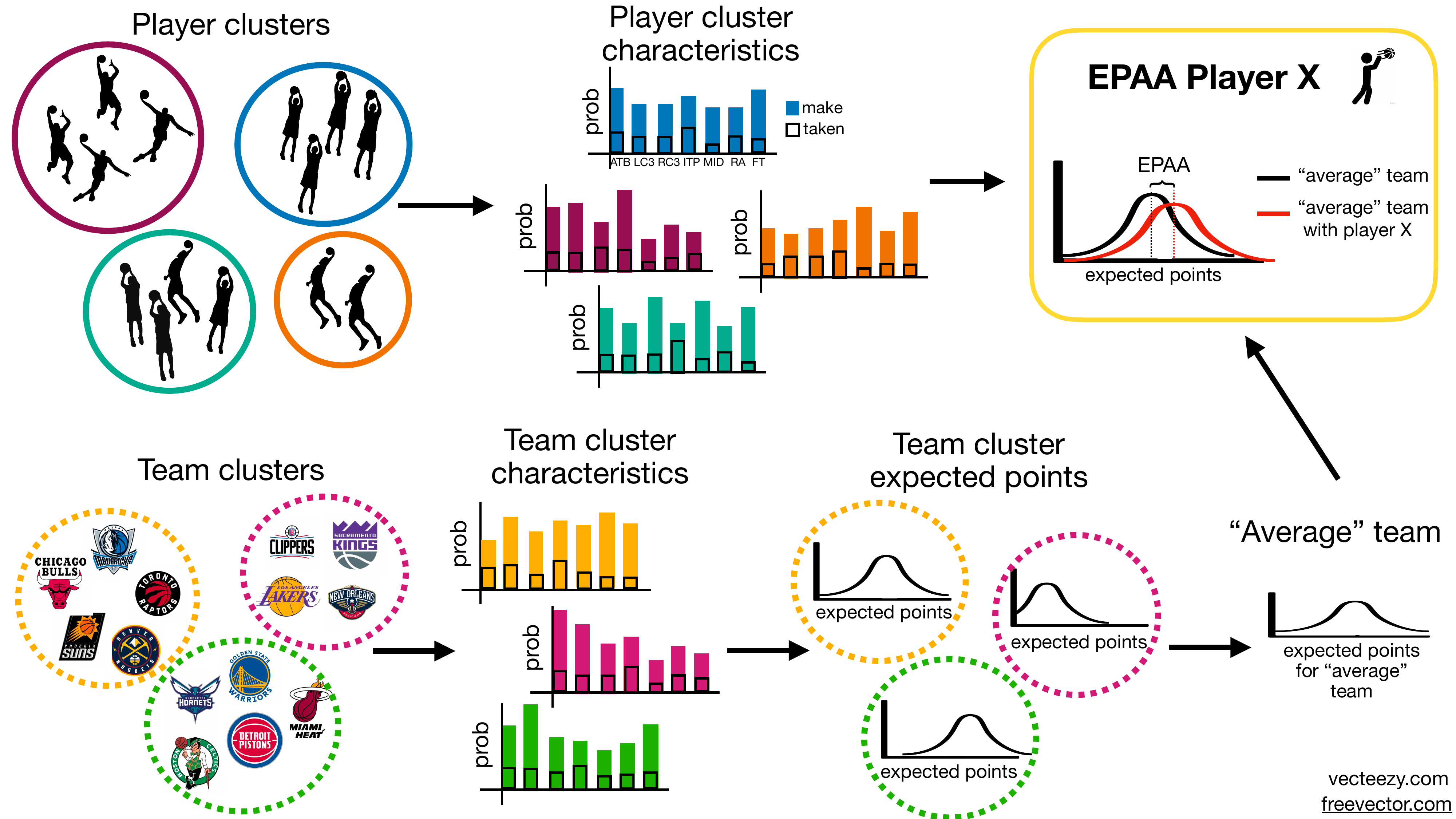}
\centering
\caption{A visual representation of the analytical pipeline used to the compute the EPAA. Along the top, we illustrate the clustering players based on their shot taking and shot making data (top left) using Bayesian hierarchical modeling. We then summarize the player clusters using the posterior distribution of the shot taking probabilities and the corresponding make probability for each region (top middle). Simultaneously, we use similar modeling framework to cluster teams (bottom left) and estimate team cluster characteristics. From the team cluster shot making/taking probabilities, we derive the posterior distribution of expected points scored in a season (bottom middle). We then average the cluster-specific expected point distributions to obtain the distribution of expected points for an average team (bottom right). Finally, to obtain the EPAA of player X (top right), we use the posterior distribution of the player cluster memberships, and the corresponding cluster characteristics, to compute the expected points of the average team if we substitute in player X. We compare this distribution (shown in red in the top right figure) to the expected point distribution for the average team and define the EPAA as the difference between the means of these distributions.}
\label{fig:bailey-fig}
\end{figure}

We perform both a team- and player-level analysis, whereby we consider the location of shots taken on the basketball court and the shot making propensities for each player or team.  Our work is similar to that of \citet{jiao_21} in that we both use Bayesian hierarchical clustering of NBA players (and teams in our case). A key difference, however, is that we utilize discretized shot locations rather than the specific $(x, y)$ coordinates on the court as the basis of our model. We do this because there are different point values associated with different regions on the court, e.g. a (made) left corner three is worth three points whereas a basket in the paint is only worth two. Spatial modeling in the Euclidean plane does not capture these nuances. Furthermore, we are less interested in the resulting cluster assignments of players (and teams). We agree with \cite{jiao_21} that players and teams have shooting commonalities, but we use these estimated clusters and their associated shooting characteristics to examine the impact of a specific player relative to an "average" player. We illustrate our data and analysis pipeline, and highlight its novelty, in Figure \ref{fig:bailey-fig}.

By using the discretized shot locations, we induce a multinomial distribution of shot selection across the court parameterized by the total number of shots taken and the propensity to take shots in the various regions. We then apply a Bayesian hierarchical modeling framework to our multinomial data to identify team or player clusters based on both the shot taking and shot making patterns.
We use the results of this team-based clustering model to derive a team-specific metric called \textit{expected points} (EP) to compare teams within and across seasons. Next, we define a player-specific metric, \textit{expected points above average} (EPAA), by combining the results of both team and player clustering using the propose Bayesian Hierarchical framework. EPAA offers a new way to evaluate players by measuring their impact on an ``average'' NBA team. See Section \ref{sssec:epaa} for more details related to how we define ``average'' in this research.

The Bayesian model employed in our analysis has been studied extensively in the past \citep[see, for example,][]{nandram1998bayesian}. As such, the novelty of our work lies in the analytical pipeline described in Figure \ref{fig:bailey-fig}. Using posterior sampling, an important deliverable of our approach is the creation of new basketball player and team evaluation tools. %For additional references on Bayesian hierarchical modeling and Bayesian statistics in general, see \citet{banerjee2003hierarchical} and \citet{gelmanbayesian}, respectively.

%We stress that the emphasis of this paper is not on the Bayesian model {\em per se}, as this model has been studied extensively in the past \citep[see, for example,][]{nandram1998bayesian}. Rather, the novelty of this work lies in the entire analytical pipeline as described in Figure \ref{fig:bailey-fig} and the use of posterior sampling to create new basketball player and team evaluation tools. For additional references on Bayesian hierarchical modeling and Bayesian statistics in general, see \citet{banerjee2003hierarchical} and \citet{gelmanbayesian}, respectively.

The remainder of the paper is organized as follows. The next section  describes the data used for the analyses. In Section 3 we describe the statistical model and method of inference. The results of our analysis are summarized in Section 4. Section 5 concludes with a discussion surrounding our main results and directions of future work.

\section{Data}
\label{sec:data}

The R software \citep{r-man} package \textit{nbastatR} \citep{nbastatr} provides access to an immense amount of historical data via the NBA's Application Programming Interface. For this project we obtained field goal and free throw data for every regular season game from the 2008-2009 season through the 2020-2021 season. Included in this data are the shot location on the court given in both $(x,y)$ coordinates and region \footnote{Region of the court is a variable returned by the team\_shot() function in the nbastatR R package.} (e.g, in the paint, corner three, etc.), the outcome of the shot (make or miss), and the team and player taking the shot for every attempted field goal and free throw. This dataset consists of over 2.6 million shot attempts spanning the 13 seasons.

\begin{figure}
\includegraphics[trim=0 500 0 0, width=5.5in]{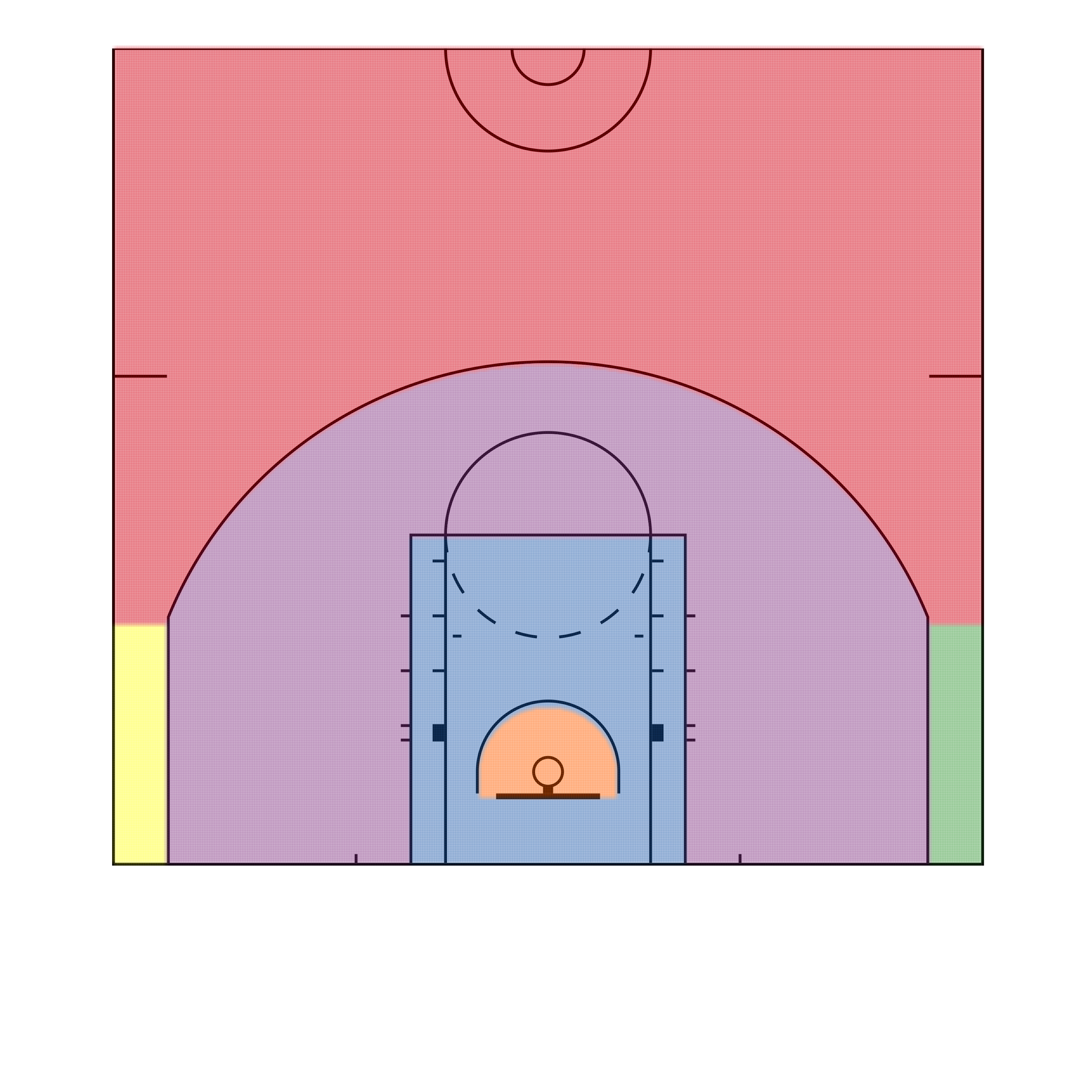}
\centering
\caption{Offensive regions on the basketball court as defined by the National Basketball Association (NBA). The regions map to their respective color as above the break three (ATB; in red), midrange (MID; purple), in the paint (ITP; blue), right corner three (RC3; yellow), left corner three (LC3; green), and restricted area (RA; orange). Note that the free throw line is the horizontal line between the solid and dashed semi-circles separating the paint (blue) and midrange (purple) regions.}
\label{fig:court_w_regions_sporty}
\end{figure}

In this analysis we are interested in examining shots taken by each team (or player) in seven offensive regions of the court (see Figure \ref{fig:court_w_regions_sporty}). The regions are defined as above the break (ATB), left corner three (LC3), right corner three (RC3), in the paint (ITP), midrange (MID), restricted area (RA), and the free throw line (FT). The first three regions are three-point shots, the next three regions are two-point shots, and FTs are worth one point. Backcourt shots (i.e., shots taken from the defensive end of the court) are not included. For each team (or player) and season, we compute the total number of shots taken and the outcome of those shots in each of the seven regions.

\renewcommand{\baselinestretch}{1}

\begin{figure}
\includegraphics[trim=0 200 0 0, width=5.5in]{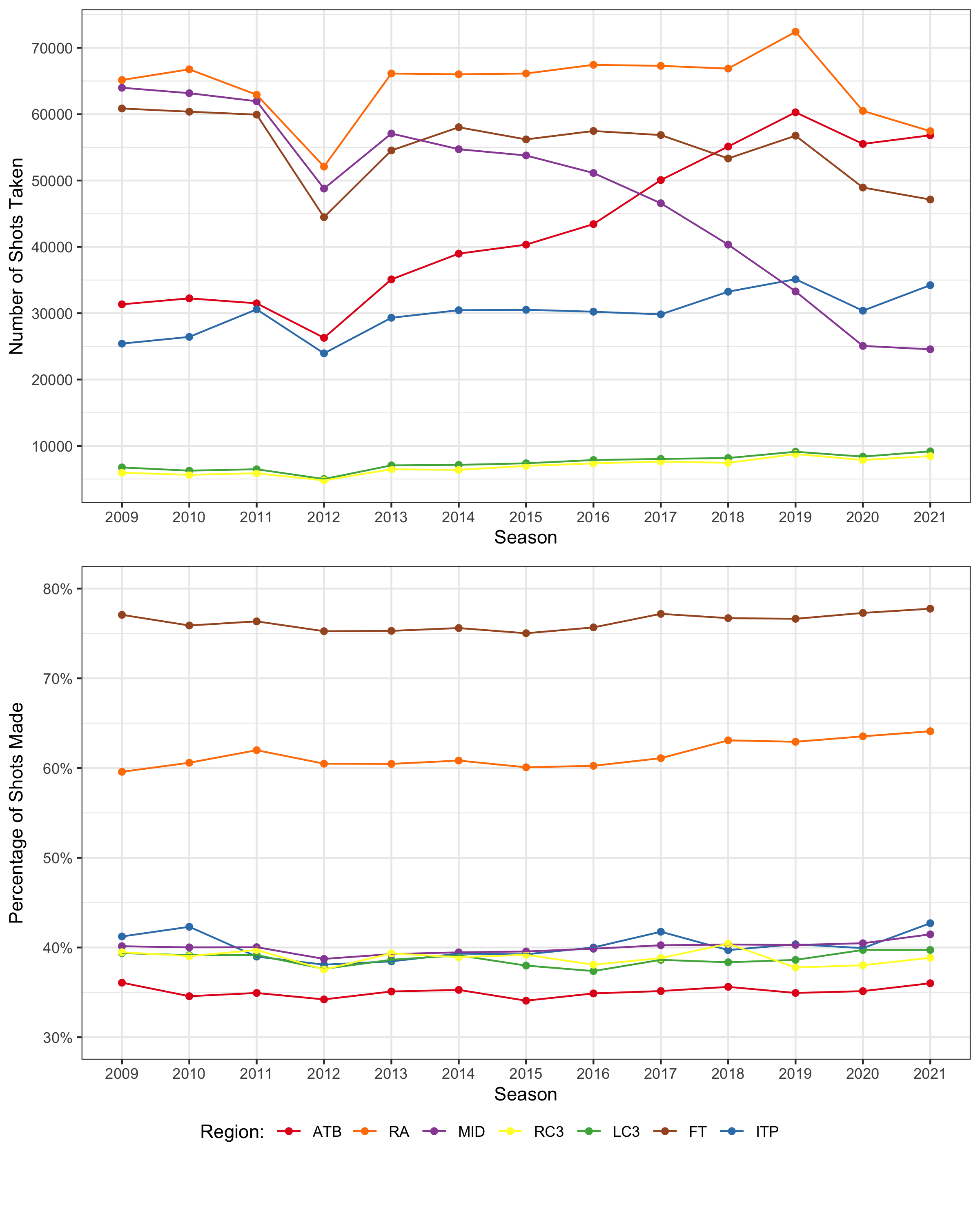}
\centering
\caption{A summarization of total shots taken (top) and percentage of shots made (bottom) across thirteen seasons in the NBA (2008-09 through 2020-21 seasons) in each of the seven NBA-defined regions on the court: above the break three (ATB), free throw (FT), in the paint (ITP), left corner three (LC3), mid-range (MID), restricted area (RA), and right corner three (RC3). Note: the significant drop in the number of shots taken during the 2011-12 season is due to that season's five-month lockout in which the regular season amounted to 66 games per team as opposed to the usual 82. The 2019-20 (varying number of games played per team) and 2020-21 (72 games per team) seasons were shortened due to the COVID-19 pandemic.}
\label{fig:all_shots}
\end{figure}

Figure \ref{fig:all_shots} (top) shows the number of shots taken by region for each season in our sample aggregated across all teams. In this and subsequent figures, the year in season refers to the year in which the playoffs were played, e.g. 2021 is actually the 2020-21 season. As is clearly evident in this figure, the shift away from the mid-range (MID) shot coincides with the increase in above the break (ATB) three point shots. This change started in the 2012/13 seasons with the number of attempted ATB three pointers peaking in the last complete season before the COVID-19 pandemic began. The league-wide shooting percentages across the various regions and seasons (Figure \ref{fig:all_shots} bottom) remain fairly consistent with a slight increase in the restricted area (RA) over time.

While these league-wide trends are easily observable, our interest is primarily in how teams and individual players behave. In particular, we are interested in identifying team- and individual-based trends and in deriving novel player-specific shooting evaluation metrics.

\section{Statistical Model and Inference}

\subsection{Bayesian Hierarchical Model Specification}
\label{sec:model}

We propose a Bayesian hierarchical model to cluster individual players and teams based on shot selection and shot making profiles. The model enables inference with regard to cluster membership as well as shot selection and accuracy cluster profiles. We first outline the approach for the team-level analysis, noting that the player-level analysis is analogous.

Let ``team'' in what follows denote a unique franchise and season combination, e.g. the Denver Nuggets in the 2020-21 season is a ``team''. To begin, the offensive half court is partitioned into the $K=7$ disjoint regions depicted in Figure \ref{fig:court_w_regions_sporty}. For each team $i = 1, \dots, I$ and region $k=1, \dots, K$, we define $N_i^{k}$ to be the number of shot attempts and $M_i^k$ to be the number of shot makes during the given season.
Let $N_i = \sum_{k} N_i^k$ and $M_i = \sum_k M_i^k$ represent the total number of shot attempts and makes, respectively, for team $i$ across all regions.

We assume each team belongs to a latent shot selection cluster, $w_i$, and a latent accuracy cluster, $z_i$. Let $L$ and $J$ denote the number of latent shot selection and shot accuracy clusters, respectively, such that $w_i \in \{1, \dots, L\}$ and $z_i \in \{1, \dots, J\}$ for the $i^{th}$ team. Note that both $L$ and $J$ are specified {\em a priori}.

For generic shot selection cluster $w$, define $p_w^k$ to be the probability of a shot being taken from region $k$, where $\sum_{k=1}^K p_w^k = 1$.
Similarly, for generic shot accuracy cluster $z$, let $q_z^k \in [0,1]$ denote the accuracy probability for region $k$. Note that the sum of the individual shot make probabilities for each region have no sum restriction. Given the $i^{th}$ team's shot selection cluster membership, $w_i$, we model the vector of shot counts for team $i$ across the $K$ regions using a Multinomial distribution:
\begin{equation}
\label{eqn:multinomial}
(N_i^1, \dots, N_i^K)|w_i \sim \text{Multinomial}\left(N_i, (p_{w_i}^1, \dots, p_{w_i}^K)\right).
\end{equation}
Next, conditional on the team's shot accuracy cluster membership, $z_i$, the number of shots made in each region $k$ follows a Binomial distribution:
\begin{equation}
\label{eqn:binomial}
M_i^k |z_i \sim \text{Binomial}(N_i^k,q^k_{z_{i}}).
\end{equation}

Within the Bayesian framework we assign probability distributions to the parameters $p_w^k$ and $q_z^k$ for all clusters $w$ and $z$.
For $w \in \{1, \dots, L\}$, each length-$K$ vector of shot selection parameters is modeled independently as
\begin{equation}
(p_w^1, \dots, p_w^K) \sim \text{Dirichlet}(\alpha, \dots, \alpha).
\end{equation}
This choice of prior assigns equal prior probability to each region $k \in \{1, \dots, K\}$ for each shot attempt cluster.  For all shot making clusters $z \in \{1, \dots, J\}$ and regions $k \in \{1, \dots, K\}$, the shot make probability, $q_z^k$, is modeled with an independent and non-informative Beta prior,
$q_z^k \sim \text{Beta}(1,1).$

To fully specify the clustering process, let $\pi_w$ denote the probability of a team being in shot attempt cluster $w$ and $\theta_z$ denote the probability of a team being in accuracy cluster $z$. Necessarily, $\sum_{l=1}^L \pi_l = \sum_{j=1}^J \theta_j = 1$.
For each team $i = 1, \dots, I$, we model the cluster membership latent variables, $w_i$ and $z_i$, as
\begin{equation}
\begin{split}
w_i &\sim \text{Multinomial}\left(1, (\pi_1, \dots, \pi_L)\right),\\
z_i &\sim \text{Multinomial}\left(1, (\theta_1, \dots, \theta_J)\right).
\end{split}
\end{equation}
Cluster membership probability vectors are further modeled using independent Dirichlet distributions. We assume equal prior probability for all clusters where
\begin{equation}
\begin{split}
(\pi_1, \dots, \pi_L) &\sim \text{Dirichlet}(\beta, \dots, \beta),\\
(\theta_1, \dots, \theta_J) &\sim \text{Dirichlet}(\gamma, \dots, \gamma).
\end{split}
\end{equation}

We use a Gibbs sampling algorithm to sample from the joint posterior distribution of the parameters given team shot counts $\boldsymbol{N}=\big\{\{N_i^k\}_{k=1}^K\big\}_{i=1}^I$ and team shot makes $\boldsymbol{M}=\big\{\{M_i^k\}_{k=1}^K\big\}_{i=1}^I$. The basic mechanics involve sampling sequentially from a series of full conditional distributions to obtain a posterior draw from each parameter in the set $\boldsymbol{\omega} = \{\boldsymbol{w}, \boldsymbol{z}, \boldsymbol{p}, \boldsymbol{q}, \boldsymbol{\pi}, \boldsymbol{\theta}\}$, where $\boldsymbol{p}$ is a length-$(K\times L)$ vector of shot selection parameters (see  \eqref{eqn:multinomial}), $\boldsymbol{q}$ is a length-$(K\times J)$ vector of shot accuracy parameters (see \eqref{eqn:binomial}), $\boldsymbol{w}$ is a length-$I$ vector of shot selection cluster memberships, $\boldsymbol{z}$ is a length-$I$ vector of shot accuracy cluster memberships, and $\boldsymbol{\pi}=(\pi_1, \dots, \pi_L)^\prime$ and $\boldsymbol{\theta}=(\theta_1, \dots, \theta_J)^\prime$ are the vectors of cluster membership probabilities. See \citet{franzen2006bayesian} for a general description of Gibbs sampling in the Bayesian mixture model context.

\subsection{Inference for NBA Analysis}
\label{sec:methods}
Model inference for the NBA analysis includes both team-level and player-level metrics. Team-level inference includes posterior estimates of \emph{expected points} (EP), defined as the predicted number of points for a specified number of shots. This quantity is computed based on the team's posterior shot taking and shot making cluster membership probabilities and the underlying cluster profiles, i.e. cluster shot taking and making probabilities per region. The total number of shot attempts is held constant to achieve a normalized comparison of expected points based on shot selection and shot making abilities across teams. Naturally, teams do not take the same number of shots within or across seasons. We fix this number so that any observed differences in expected points or expected points above average is due to the team's shooting characteristics and not the raw number of shots taken. Similar comparisons are made for individuals based on a player-level analysis for a given number of shots. The player- and team-based clustering serves as the basis for our novel, basketball-specific metric, \emph{expected points above average} (EPAA), to evaluate a player's contribution (positive or negative) relative to the performance of an ``average'' player. The EP and EPAA metrics are outlined in detail in the following subsections.

\subsubsection{Expected Points}
\label{sec:ep}

One of the primary benefits of employing a Bayesian hierarchical framework in this NBA setting is that it enables estimation and uncertainty quantification based on the full posterior distributions of team- and player-based characteristics. Here, we focus on deriving the posterior distribution of a team's expected points, which can be obtained for an actual or hypothetical team.

Suppose a team takes $\tilde{N}$ shots. The goal is to estimate total points (either per game, season total, etc.) based on the hypothetical team's characteristics using the Bayesian hierarchical model from Section \ref{sec:model}. This requires deriving the posterior predictive distribution of shots made across each region, $\tilde{\boldsymbol{M}}$, given hypothetical shot count $\tilde{N}$ and the observed data, $\boldsymbol{N}$ and $\boldsymbol{M}$. Formally, this can be written

$$
 f(\tilde{\boldsymbol{M}} | \tilde{N}, \boldsymbol{N}, \boldsymbol{M}) = \int_{\boldsymbol{\Omega}} \sum_{\tilde{\boldsymbol{N}}} f(\tilde{\boldsymbol{M}} | \tilde{\boldsymbol{N}}, \boldsymbol{\omega}) f(\tilde{\boldsymbol{N}} | \tilde{N}, \boldsymbol{\omega}) f(\boldsymbol{\omega} | \boldsymbol{N}, \boldsymbol{M}) \textrm{d}\boldsymbol{\omega}.
$$

Given our posterior draws of the parameters we can easy sample from this posterior predictive distribution. Each draw from the posterior distribution leads to the posterior probability of cluster membership (shot selection and accuracy) using Bayes' theorem conditioned on the appropriate parameters. We then sample the number of shots per court region from a cluster-specific shot-selection Multinomial distribution. Within each region, we sample the number of shots made in each region based on the cluster-specific accuracy Binomial distributions. Total points for the posterior predictive draw is computed as the weighted sum of shots made across the seven regions where the weights are equal to the point value for each region. 

\subsubsection{Expected Points Above Average}
\label{sssec:epaa}

Let $\tilde{N}^i$ denote a hypothetical number of shots taken by player $i$. To compute EPAA, we compute the difference in expected points based on $\tilde{N}^i$ shots had they been taken by player $i$ versus if they were taken by an ``average'' team. In both cases (player and average team), we sample from the posterior predictive distributions as described in Section 3.2.1 given $\tilde{N}^i$ total shots. 

Specifically, for player $i$, we sample from the posterior predictive distribution given $\tilde{N}^i$ shots based on the player-level model. We then sample from the posterior predictive distribution given $\tilde{N}^i$ shots using the team-level model. That is, we randomly select a team and use that team's characteristics for each sample from the team-level model to construct an "average" team's distribution of points given $\tilde{N}^i$ shots. The distribution of the difference in point totals between the player and the average team then represents the impact of player $i$. 

It should be noted that we did not compute the "average" team based on randomly sampling individual players. This approach would have led to biased results as our individual player model was fitted only to the top 100 shot takers during the 2020-2021 season.

%\section{Data Analysis}
\section{NBA Player and Team Data Analyses}
\label{sec:apps}

We fix the model cluster specification in the following manner. We set $L$ and $J$, the number of shot selection and shot accuracy clusters, respectively, to $L=J=20$, which allows for a total of 400 shot selection and shot accuracy cluster combinations for both players and teams. Furthermore, we set the nuisance parameters at $\alpha = \beta = \gamma = 5$ in this analysis. We examined the similarity of the resulting clusters with respect to the estimated shot taking and making probabilities in Section S1 of the online supplement \citep{elmoresup}.

Through a sensitivity analysis, we investigate additional choices of $L$ and $J$, including $L,J \in \{10, 20, 30\}$ as well as other specifications of the parameters $\alpha = \beta = \gamma$. Based on our exploration, there were only minor differences in expected points per game and EPAA, respectively, when increasing $L$ or $J$ beyond 20 (Figures 3 and 4 and Figures 7 and 8 of the online supplement). We did not detect much change in the expected points per game or EPAA when varying $\alpha$, $\beta$, and $\gamma$ (Figures 5, 6, 9, and 10 in the online supplement). Detailed results of this sensitivity analysis are provided in Section S2 of the online supplement.

%(e.g., $L = J = 10, 20,$ and $30$; $L=10$ and $J=30$; and $L=30$, $J=10$) 

%As part of this analysis we conducted a sensitivity study with respect to the choice of $L$ and $J$, and the nuisance parameters, $\alpha$, $\beta$, and $\gamma$.  The primary conclusion of the sensitivity study is that, although we observe some team-to-team and player-to-player variation in the end results, the general trends are only marginally affected.

Each team's initial shot selection and shot accuracy cluster memberships ($\boldsymbol{w}$ and $\boldsymbol{z}$, respectively) are determined using k-means clustering \citep{james2013introduction} based on their observed shot frequencies and makes within the pre-defined regions on the court. The cluster profile parameters $\boldsymbol{p}$ and $\boldsymbol{q}$ were set to the mean values of the teams associated with respective cluster, and the cluster membership probability vectors $\boldsymbol{\pi}$ and $\boldsymbol{\theta}$ were set to the empirical fraction of teams assigned to each cluster through k-means.

\subsection{MCMC Sampling}
\label{sec:mcmc_samples}

We ran the sampling algorithm for 10,000 iterations. No issues of convergence were detected based on (1) a visual examination of the trace plots and (2) effective sample size calculations for the posterior samples. Additional details related to convergence are presented in Section S3 of the online supplement. The first 3,000 iterations were discarded as burn-in and the remaining 7,000 posterior samples were used for inferential investigation of NBA teams and players. The team- and player-level comparisons based on expected points and expected points above average are reported in Section \ref{sec:ep_epaa}. 

\subsection{Expected Points and Expected Points Above Average}
\label{sec:ep_epaa}

The Bayesian hierarchical methodology enables posterior inference with regard to expected points and expected points above average. To begin, let $\tilde{N} = 8000$, be the total number of shots taken by each team in our analysis. This number represents approximately the average number of total shots taken (field goals and free throws) per team during the 2020-2021 season. Using a common number of shots provides an equal basis for comparing teams based on their shot taking and shot making abilities, and not simply on the total number of shots a team takes during a season. In fixing $\tilde{N}$, the predicted point totals need not mimic the actual points scored by a team during the season due to each team taking more or less than 8000 total shots.

\begin{figure}
\includegraphics[width=5.5in]{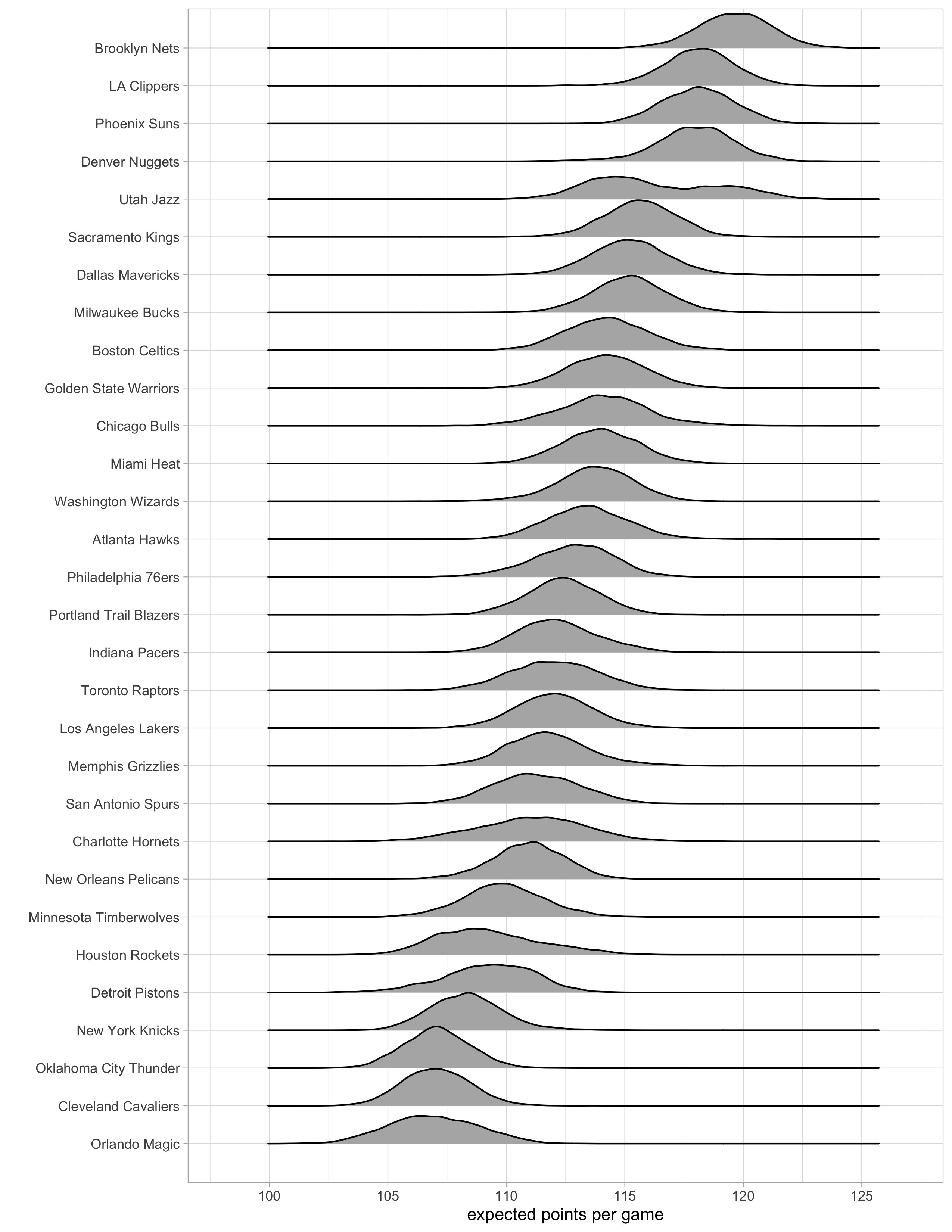}
\centering
\caption{Posterior predictive distributions of expected points per game for 2020-21 season for each NBA team. Predictive distributions were normalized such that the results are based on each team taking 8000 shots during the season. Differences across teams are due to differences in membership of shot selection and shot accuracy clusters.}
\label{fig:exp_pts}
\end{figure}

We compute the posterior predictive distribution of expected points for each team in the 2020-21 season based on the approach outlined in Section \ref{sec:methods}. The predictive distributions of expected points are displayed in Figure \ref{fig:exp_pts} for each team. Based on our model and historical performance we would expect the top scoring team to be the 2020-21 Brooklyn Nets with an estimated total of 120 points per game based on 8000 shots over the season. The 2020-21 Nets were led by Kevin Durant, James Harden, and Kyrie Irving, three players known for their scoring prowess. The LA Clippers, Phoenix Suns, and Denver Nuggets round out the top four teams. The bottom four teams are the New York Knicks, Oklahoma City Thunder, Cleveland Cavaliers, and the Orlando Magic.

To evaluate and compare individual players, we next examine EPAA as outlined in subsection \ref{sec:methods}. Recall that this requires posterior inference from both the team- and individual-based models.  Thus, we refit the Bayesian clustering model to the top 100 shot takers for the 2020-2021 seasons. We only used this subset of all players, i.e. the high-volume shooters, to investigate the EPAA as a player-specific metric. We will briefly describe three applications in the following paragraphs.

\begin{figure}
\centering
\includegraphics[width=5.5in]{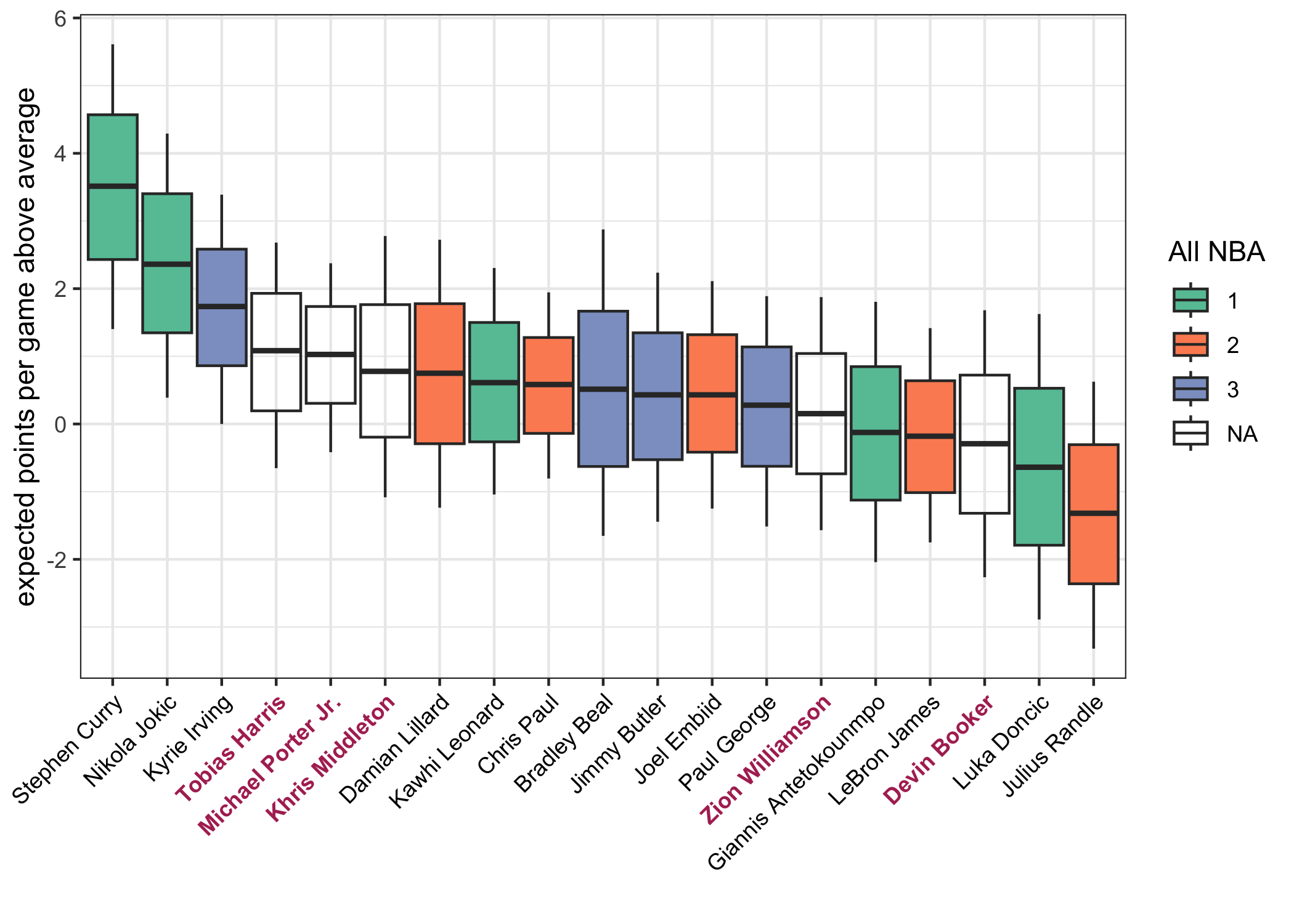}
\caption{Posterior distributions of the expected points above average per game for nineteen players from the 2020-21 season. The black horizontal line represents the posterior median, the box contains the 80\% credible interval and whiskers delineate the 95\% credible interval for each player. The players include fourteen of the fifteen All NBA (1st, 2nd, and 3rd Team) and five players having high EPAA that were not selected for the All NBA team. Rudy Gobert is omitted due to not having a high enough shot volume. Players are sorted from greatest (left) to least (right) posterior predictive median value. The players highlighted in maroon did not make the All NBA teams.}
\label{fig:epaa_all_nba}
\end{figure}

Figure \ref{fig:epaa_all_nba} shows posterior mean and 95\% credible intervals estimates of EPAA for nineteen players from the 2020-21 season. The list of players includes fourteen of the All NBA first, second, and third teams (Rudy Gobert is omitted due to a lack of shots) as well as five additional players with high EPAA who were not selected for an All NBA team (highlighted in maroon). Note, the All-NBA teams are a recognition of the top 15 (5 per team) players as voted by members of the sports media community. The five players in maroon could be considered All NBA ``snubs'', or players whose shooting success could conceivably have warranted them a spot on All NBA, but were omitted. This analysis shows that traditionally high-volume and accurate three-point shooters tend to have the largest impact above an average team taking the same number of shots with Stephen Curry ranking first. Note also that Tobias Harris, Michael Porter Jr., and Khris Middleton rank four, five, and six among these players and were not selected for All NBA. Interestingly, however, the 2020-21 NBA Most Valuable Player, Nikola Joki\'{c}, who plays center, is ranked second in terms of mean EPAA. This analysis indicates EPAA can be used as an additional objective measure for a player's All NBA portfolio.

\begin{figure}
\includegraphics[width=5.5in]{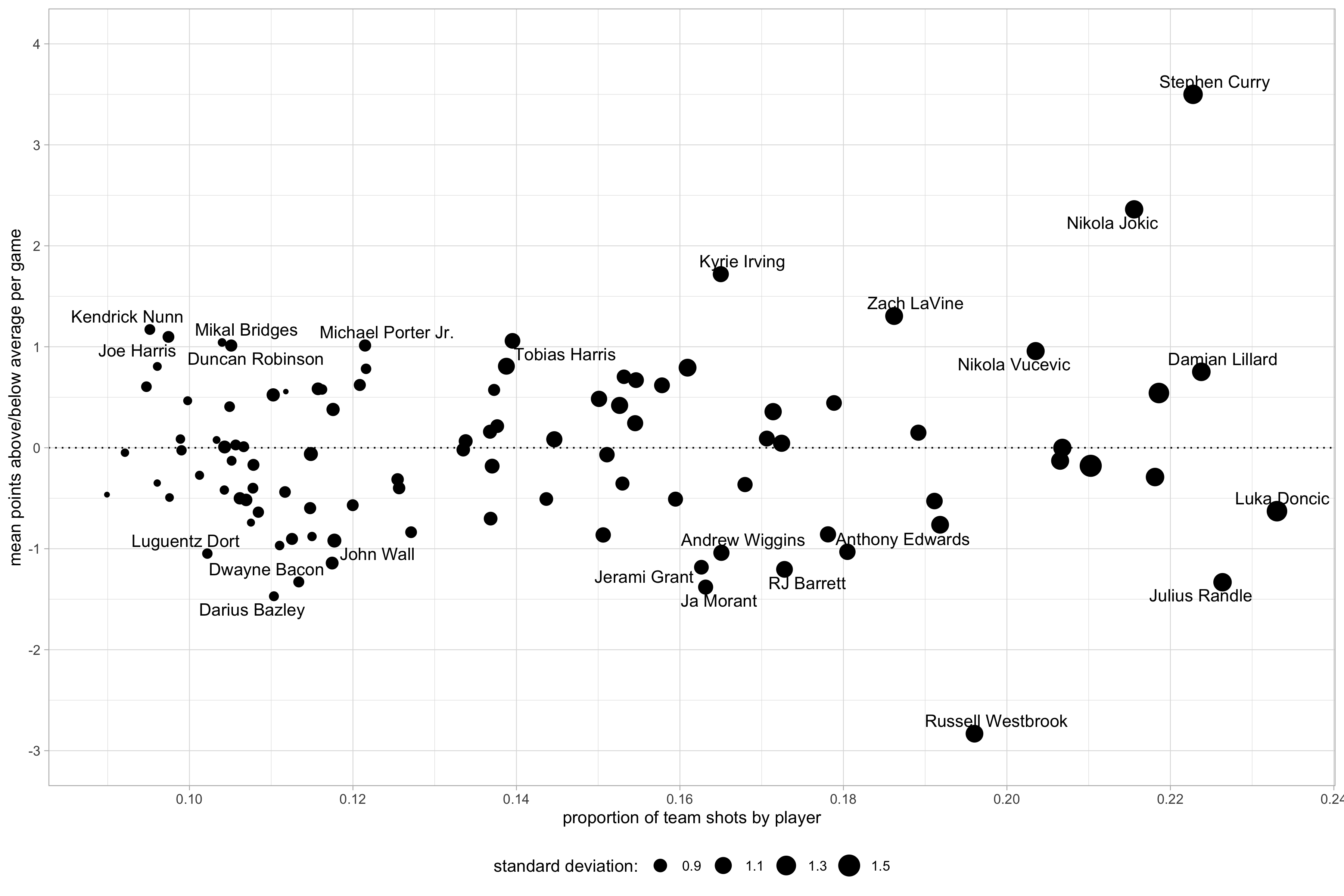}
\centering
\caption{Posterior predictive mean points above average versus the proportion of shots taken by a player during the 2020-21 NBA season. Several players on the periphery are labeled for reference. Points are sized based on the standard deviation of points above average.}
\label{fig:paa_by_prop}
\end{figure}

It may be tempting to attribute the ordering in Figure \ref{fig:epaa_all_nba} to the proportion of shots taken by each player relative to their team during that season. However, there is no meaningful correlation between a player's EPAA and the proportion of the team's shots taken by the player (see Figure \ref{fig:paa_by_prop}). In fact, a diagnostic such as this could be used to advocate on behalf of players that are not shooting at a very high volume relative to their team and yet still have a high EPAA (upper left), e.g. Joe Harris, Kendrick Nunn, Mikal Bridges, and Duncan Robinson.

\begin{figure}
\includegraphics[width=5.5in]{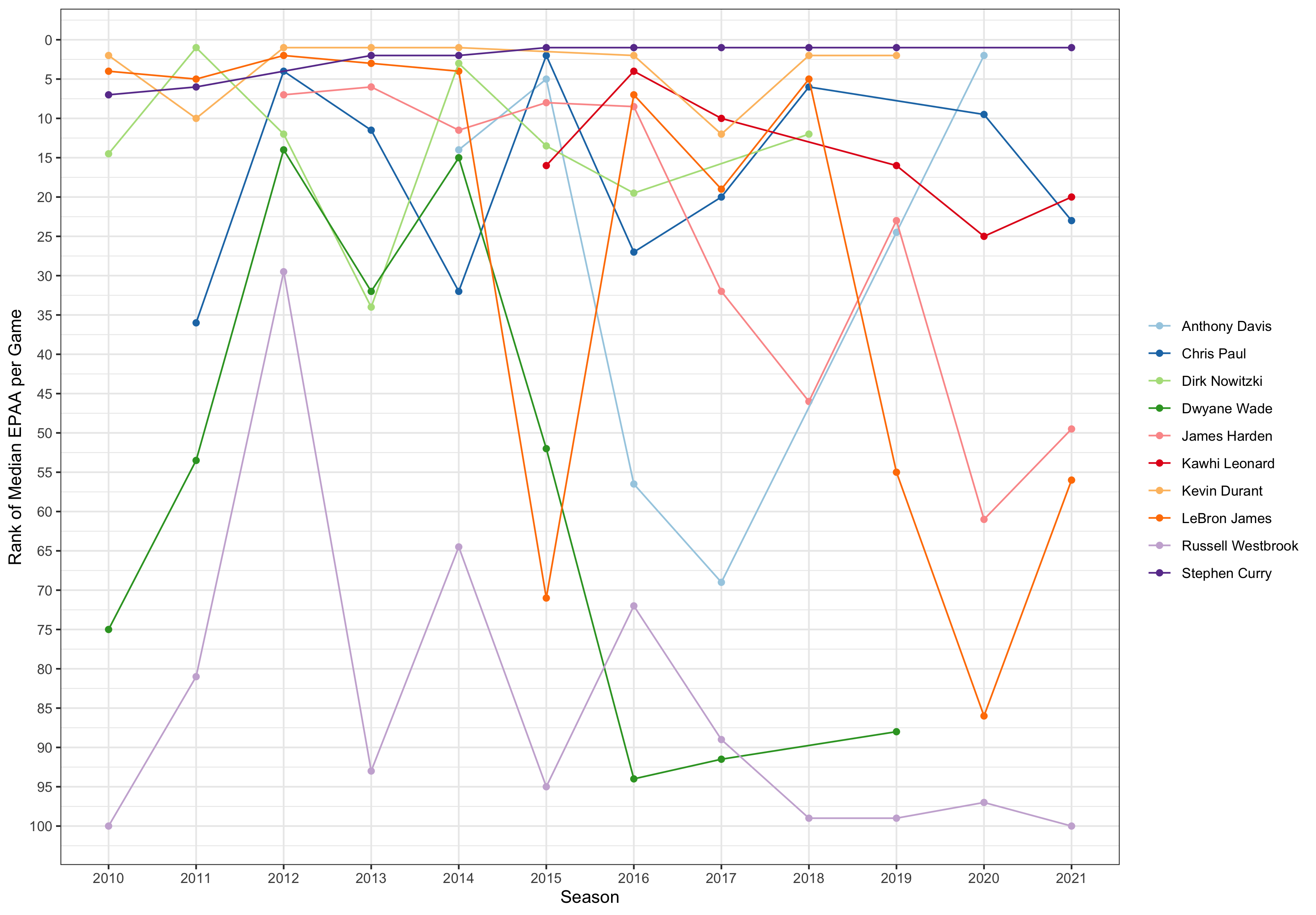}
\centering
\caption{Ranking by year of ten of the top eleven players (according to ChatGPT) from the 2010s in terms of their mean EPAA. Note that we excluded Kobe Bryant due to having only qualified for three years in this analysis. }
\label{fig:ranks_epaa}
\end{figure}

Next, we investigate the ranking within year of the posterior predictive mean EPAA of ten of the best players from the 2010s across each season for which they qualified (i.e., one of the top 100 shot takers per season). The ten players were chosen based on a query to ChatGPT asking for the top fifteen players of the 2010s. These players are the top eleven minus Kobe Bryant, who only qualified for three years in the 2010s. The results are given in Figure \ref{fig:ranks_epaa}. It is interesting to note that Dirk Nowitzki ranked in the top 20 (except 2013) even though he was nearing the end of his career in the 2010s. Furthermore, it shows how dominant Stephen Curry and Kevin Durant have been during this time period. We also clearly see the peak and decline of Russell Westbrook and the gradual decline of Dwyane Wade as he ended his career.

\begin{figure}%
    \centering
    {{\includegraphics[width=2.5in]{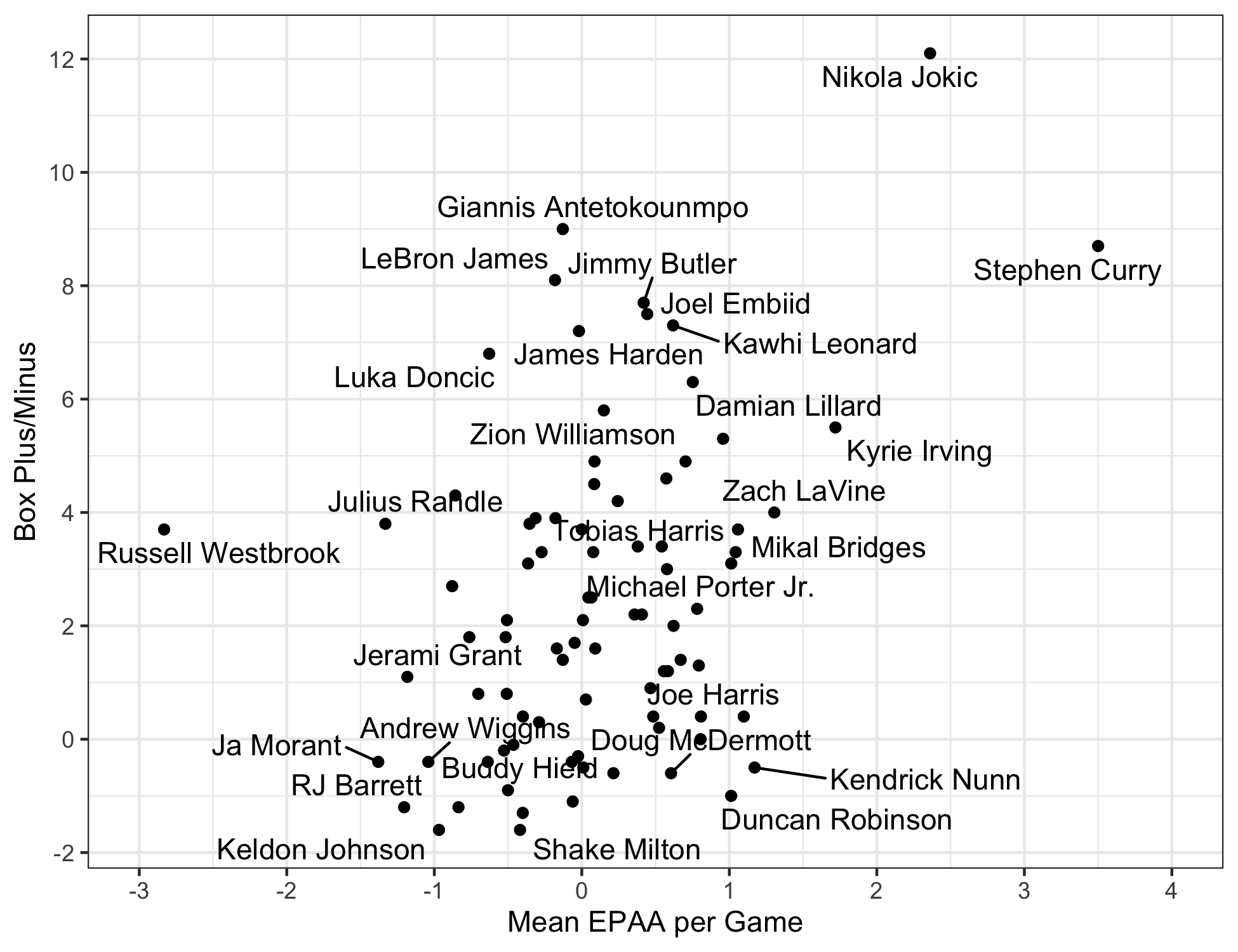} }}%
    \qquad
    {{\includegraphics[width=2.5in]{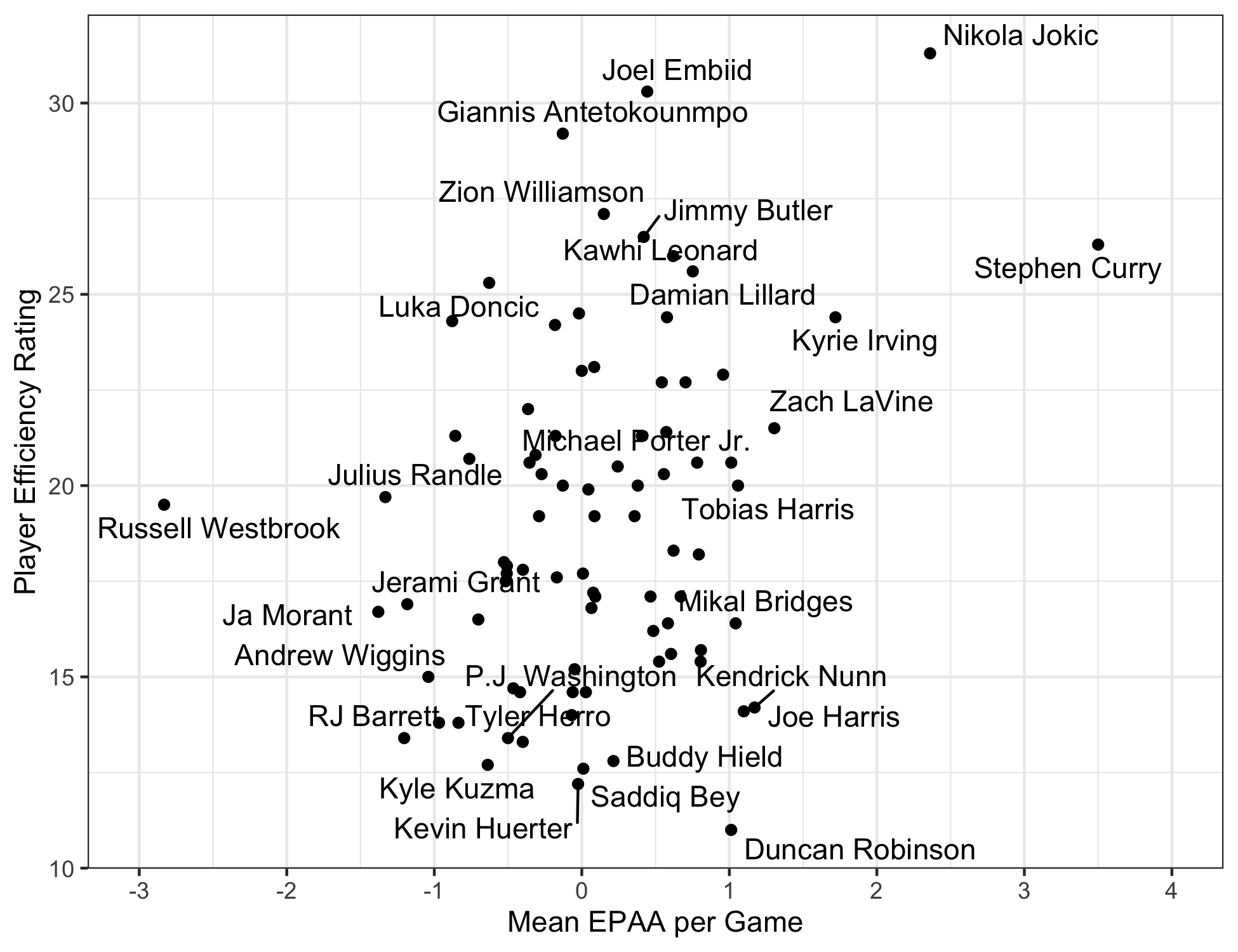} }}%
    \caption{The two panels show the mean EPAA for the top one hundred shot takers in the 2020-21 NBA season plotted against: (left) the player's Box Plus/Minus value and (right) the player's Player Efficiency Rating for the same season.}%
    \label{fig:bpm_per}%
\end{figure}

Finally, we investigate how the EPAA compares to both Player Efficiency Rating and Box Plus/Minus. In Figure \ref{fig:bpm_per} we see the overall positive association between these existing measures and EPAA. However the Pearson correlation between EPAA and PER, and EPAA and Box Plus/Minus is only 0.246 and 0.238, respectively.  Meanwhile, the correlation between PER and Box Plus/Minus is 0.915.  This demonstrates that EPAA quantifies unique aspects of player offensive efficiency not captured by existing measures.

\section{Web Application}

\begin{figure}
\includegraphics[width=5.5in]{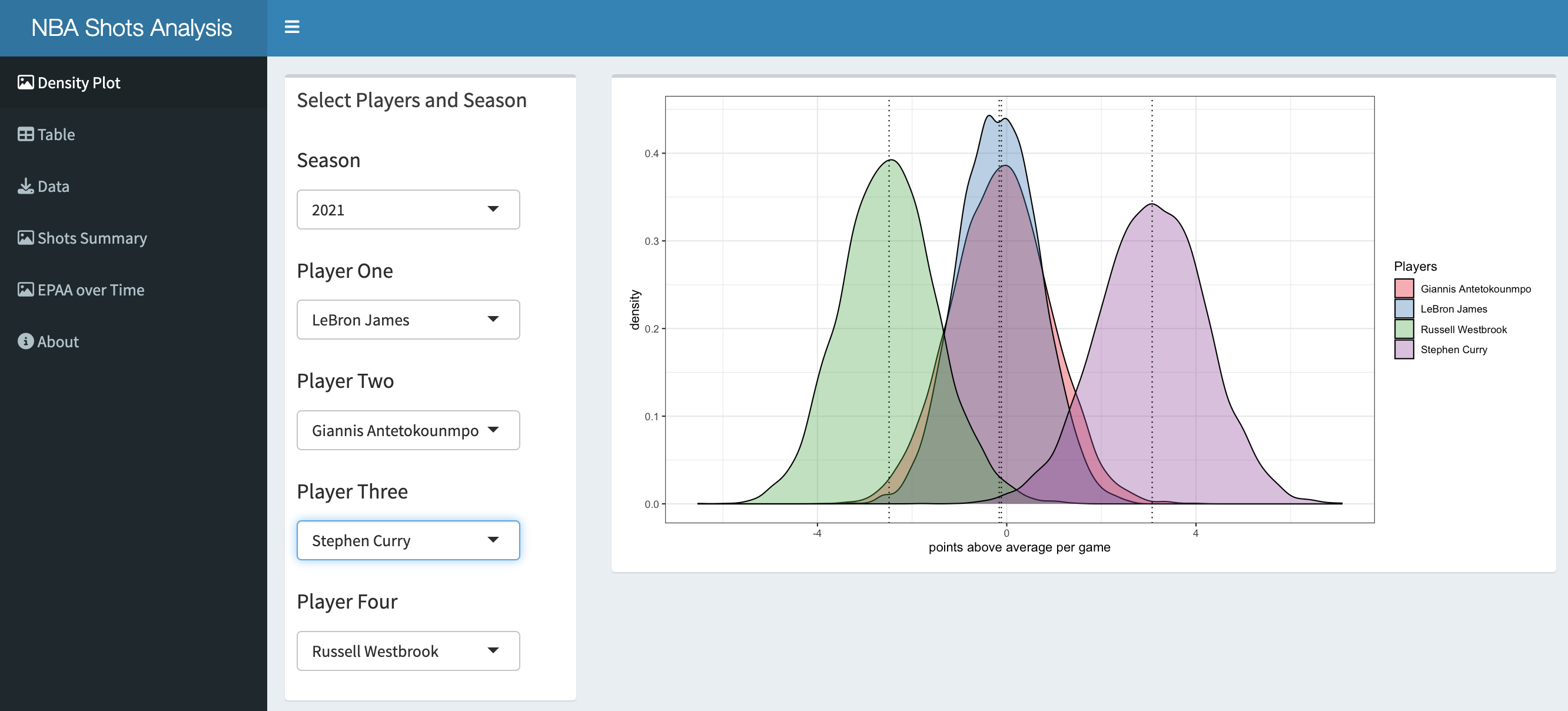}
\centering
\caption{The Shiny application showing the EPAA player-comparison density plots from the 2020-21 season for LeBron James, Giannis Antetokounmpo, Stephen Curry, and Russel Westbrook.}
\label{fig:shiny-app}
\end{figure}

To increase the utility of EPAA as a player evaluation tool, we provide a Shiny application \citep{shiny} as a means to compare the performance of players across three seasons (2017-18, 2018-19, and 2020-21). The application has five key components: (1) density plots of EPAA for up to four players per season, (2) a sortable table of all players for which EPAA was calculated, (3) a download button with access to the posterior draws for EPAA calculations, (4) an interactive scatterplot showing a team's shot taking and accuracy trends over time, and (5) EPAA over time for up to four players. The full application can be found at \citep{shiny-epaa} and the landing page is shown in Figure \ref{fig:shiny-app}. See the online supplement S5 for additional information related to the code to produce the application and the application itself.

\section{Conclusions and Discussion}
\label{sec:conclusions}

In this research we develop a Bayesian hierarchical model to cluster NBA teams and players based on their shot taking tendencies and shot making abilities. Our approach enables full probabilistic inference of important team- and player-based metrics, including expected points and a novel expected points above average metric. The methodology is highlighted using team- and player-based comparisons from the 2020-21 season. 

The NBA game has naturally evolved over time, with the most obvious change being a greater emphasis on three-point shooting in recent years. We utilized the clustering methodology described in Section \ref{sec:methods} to account for changes, as well as commonalities, in shooting characteristics over time. Through the {\em use} of posterior samples, the contribution of this work was  to develop new basketball metrics. % and {\em not} on inference with respect to estimated clusters themselves. 
Importantly, expected points and EPAA aligns with each of the three goals of sports analytics research as outlined in the introduction: evaluating individual impact, improving winning, and quantifying components of the game. In particular, EPAA is a novel method to evaluate the scoring of an individual basketball player, which enables teams to better identify players who can contribute to their success. Both metrics represent novel methods for quantifying shooting in the sport of basketball.

In addition, the EPAA metric provides a novel way to evaluate individual impact. The application could improve a team's overall quality by allowing team management to identify the best offensive player and quantifying components of the game by examining team shooting profiles over time. From a management perspective, this type of analysis would help to identify players who have similar performance metrics yet are paid at markedly different rates. 

We believe that this methodology and our new metrics, Expected Points and Expected Points Above Average, could easily be extended to other sporting domains. For example, this methodology and similar concepts could be applied to data in the National Football League (NFL) to cluster teams and potentially develop novel player-specific metrics, e.g. passing yards above average for quarterbacks. Furthermore, shooting and/or passing tendencies among soccer teams and players in the English Premier League (EPL), Major League Soccer (MLS), or other leagues are natural extensions. Likewise, a similar analysis could shed more insight into playing tendencies in the National Hockey League (NHL).

Despite these benefits of our proposed model and inference, there are a few limitations to this work. First, placing NBA (or any sport) players on an "average" team is an impossible task in practice. However, we believe, as do many practitioners, that it offers insight into a new dimension of individual player performance. Basketball is a team game, and metrics such as EPAA ignore potential synergistic effects between teammates as they relate to scoring. Additionally, under different nuisance parameter settings, we observed slight variations in the posterior distributions of expected points per game and EPAA for teams and players, respectively. Importantly, however, this had little impact on conclusions drawn from the results of our model given that the ranking of the teams and players remained fairly consistent across these specifications. % when drawing conclusions from our model. %making overall inference and conclusions and, therefore, we are comfortable with our overall results and conclusions.

There are several possible directions for extending this work. Specifically, estimating the number of both shot taking and accuracy clusters in a data-driven manner is ripe for exploration. The problem is well known in the literature on finite mixture models, see, for example, \citet{mclachlan2000finite}. Unlike the Bayesian approach offered here in which Dirichlet priors are used, the utilization of sparse finite mixture models to infer the number of clusters is becoming increasingly popular \citep{fruhwirth2019here, malsiner2016model}. A comparison of this approach with our current analytical pipeline would certainly be an interesting study.

%%%%%%%%%%%%%%%%%%%%%%%%%%%%%%%%%%%%%%%%%%%%%%
%% Supplementary Material, including data   %%
%% sets and code, should be provided in     %%
%% {supplement} environment with title      %%
%% and short description. It cannot be      %%
%% available exclusively as external link.  %%
%% All Supplementary Material must be       %%
%% available to the reader on Project       %%
%% Euclid with the published article.       %%
%%%%%%%%%%%%%%%%%%%%%%%%%%%%%%%%%%%%%%%%%%%%%%
%\begin{supplement}
%\stitle{???}
%\sdescription{???.}
%\end{supplement}

\begin{supplement}[id=suppA]
\sname{Supplement A}
\stitle{Additional Information}
\slink[doi]{TBD}
\sdatatype{.pdf}
\sdescription{Summary of our cluster simiarity study (S1), sensitivity analyses (S2), convergence of the MCMC algorithm (S3), a detailed explanation of the MCMC algorithm (S4), and references to code on Github and Shiny Application (S5).}
\end{supplement}

%% if your bibliography is in bibtex format, uncomment commands:
\newpage
\bibliographystyle{imsart-nameyear} % Style BST file
\bibliography{bibliography}       % Bibliography file (usually '*.bib')

\begin{thebibliography}{30}
% BibTex style file: imsart-nameyear.bst, 2017-11-03
% Default style options (sort=1,type=nameyear).
% Used options (sort=1,type=nameyear).

\bibitem[\protect\citeauthoryear{Baumer, Jensen and Matthews}{2015}]{openwar}
\begin{barticle}[author]
\bauthor{\bsnm{Baumer},~\bfnm{Benjamin~S.}\binits{B.~S.}},
  \bauthor{\bsnm{Jensen},~\bfnm{Shane~T.}\binits{S.~T.}} \AND
  \bauthor{\bsnm{Matthews},~\bfnm{Gregory~J.}\binits{G.~J.}}
(\byear{2015}).
\btitle{openWAR: An open source system for evaluating player performance in
  major league baseball}.
\bjournal{Journal of Quantitative Analysis in Sports}
\bvolume{11}
\bpages{69--84}.
\end{barticle}
\endbibitem

\bibitem[\protect\citeauthoryear{Bresler}{2021}]{nbastatr}
\begin{bmanual}[author]
\bauthor{\bsnm{Bresler},~\bfnm{Alex}\binits{A.}}
(\byear{2021}).
\btitle{nbastatR: R's interface to NBA data}
\bnote{R package version 0.1.1506}.
\end{bmanual}
\endbibitem

\bibitem[\protect\citeauthoryear{Chang et~al.}{2023}]{shiny}
\begin{bmanual}[author]
\bauthor{\bsnm{Chang},~\bfnm{Winston}\binits{W.}},
  \bauthor{\bsnm{Cheng},~\bfnm{Joe}\binits{J.}},
  \bauthor{\bsnm{Allaire},~\bfnm{JJ}\binits{J.}},
  \bauthor{\bsnm{Sievert},~\bfnm{Carson}\binits{C.}},
  \bauthor{\bsnm{Schloerke},~\bfnm{Barret}\binits{B.}},
  \bauthor{\bsnm{Xie},~\bfnm{Yihui}\binits{Y.}},
  \bauthor{\bsnm{Allen},~\bfnm{Jeff}\binits{J.}},
  \bauthor{\bsnm{McPherson},~\bfnm{Jonathan}\binits{J.}},
  \bauthor{\bsnm{Dipert},~\bfnm{Alan}\binits{A.}} \AND
  \bauthor{\bsnm{Borges},~\bfnm{Barbara}\binits{B.}}
(\byear{2023}).
\btitle{shiny: Web Application Framework for R}.
\end{bmanual}
\endbibitem

\bibitem[\protect\citeauthoryear{Daly-Grafstein and Bornn}{2020}]{daly}
\begin{barticle}[author]
\bauthor{\bsnm{Daly-Grafstein},~\bfnm{Daniel}\binits{D.}} \AND
  \bauthor{\bsnm{Bornn},~\bfnm{Luke}\binits{L.}}
(\byear{2020}).
\btitle{Using in-game shot trajectories to better understand defensive impact
  in the {NBA}}.
\bjournal{Journal of Sports Analytics}
\bvolume{6}
\bpages{235--242}.
\end{barticle}
\endbibitem

\bibitem[\protect\citeauthoryear{Deshpande and Jensen}{2016}]{DeshpandeJensen}
\begin{barticle}[author]
\bauthor{\bsnm{Deshpande},~\bfnm{Sameer~K.}\binits{S.~K.}} \AND
  \bauthor{\bsnm{Jensen},~\bfnm{Shane~T.}\binits{S.~T.}}
(\byear{2016}).
\btitle{Estimating an NBA player’s impact on his team’s chances of
  winning}.
\bjournal{Journal of Quantitative Analysis in Sports}
\bvolume{12}
\bpages{51--72}.
\bdoi{doi:10.1515/jqas-2015-0027}
\end{barticle}
\endbibitem

\bibitem[\protect\citeauthoryear{D'Urso, De~Giovanni and
  Vitale}{2022}]{durso22}
\begin{barticle}[author]
\bauthor{\bsnm{D'Urso},~\bfnm{Pierpaolo}\binits{P.}},
  \bauthor{\bsnm{De~Giovanni},~\bfnm{Livia}\binits{L.}} \AND
  \bauthor{\bsnm{Vitale},~\bfnm{Vincenzina}\binits{V.}}
(\byear{2022}).
\btitle{A Bayesian network to analyse basketball players’ performances: a
  multivariate copula-based approach}.
\bjournal{Annals of Operations Research}
\bvolume{325}
\bpages{419 -- 440}.
\end{barticle}
\endbibitem

\bibitem[\protect\citeauthoryear{Elmore}{2024}]{shiny-epaa}
\begin{bmisc}[author]
\bauthor{\bsnm{Elmore},~\bfnm{Ryan}\binits{R.}}
(\byear{2024}).
\btitle{{NBA} Shots Analysis}.
\bnote{\url{https://ryan-elmore.shinyapps.io/NBA-EPAA/}}.
\end{bmisc}
\endbibitem

\bibitem[\protect\citeauthoryear{Elmore and Urbaczewski}{2025}]{isar}
\begin{bbook}[author]
\bauthor{\bsnm{Elmore},~\bfnm{Ryan}\binits{R.}} \AND
  \bauthor{\bsnm{Urbaczewski},~\bfnm{Andrew}\binits{A.}}
(\byear{2025}).
\btitle{Introduction to Sports Analytics using R}.
\bpublisher{Prospect Press}.
\end{bbook}
\endbibitem

\bibitem[\protect\citeauthoryear{Franz{\'e}n}{2006}]{franzen2006bayesian}
\begin{barticle}[author]
\bauthor{\bsnm{Franz{\'e}n},~\bfnm{Jessica}\binits{J.}}
(\byear{2006}).
\btitle{Bayesian inference for a mixture model using the gibbs sampler}.
\bjournal{MResearch Report}
\bvolume{1}.
\end{barticle}
\endbibitem

\bibitem[\protect\citeauthoryear{Fr{\"u}hwirth-Schnatter and
  Malsiner-Walli}{2019}]{fruhwirth2019here}
\begin{barticle}[author]
\bauthor{\bsnm{Fr{\"u}hwirth-Schnatter},~\bfnm{Sylvia}\binits{S.}} \AND
  \bauthor{\bsnm{Malsiner-Walli},~\bfnm{Gertraud}\binits{G.}}
(\byear{2019}).
\btitle{From here to infinity: sparse finite versus Dirichlet process mixtures
  in model-based clustering}.
\bjournal{Advances in data analysis and classification}
\bvolume{13}
\bpages{33--64}.
\end{barticle}
\endbibitem

\bibitem[\protect\citeauthoryear{Gareth et~al.}{2013}]{james2013introduction}
\begin{bbook}[author]
\bauthor{\bsnm{Gareth},~\bfnm{James}\binits{J.}},
  \bauthor{\bsnm{Witten},~\bfnm{Daniela}\binits{D.}},
  \bauthor{\bsnm{Hastie},~\bfnm{Trevor}\binits{T.}} \AND
  \bauthor{\bsnm{Tibshirani},~\bfnm{Robert}\binits{R.}}
(\byear{2013}).
\btitle{{An Introduction to Statistical Learning}}
\bvolume{112}.
\bpublisher{Springer}.
\end{bbook}
\endbibitem

\bibitem[\protect\citeauthoryear{Hollinger}{2007}]{hollinger}
\begin{bmisc}[author]
\bauthor{\bsnm{Hollinger},~\bfnm{John}\binits{J.}}
(\byear{2007}).
\btitle{What is {PER}?}
\bnote{\url{https://www.espn.com/nba/columns/story?columnist=hollinger_john&id=2850240}}.
\end{bmisc}
\endbibitem

\bibitem[\protect\citeauthoryear{Hu, Yang and Xue}{2021}]{hu2021bayesian}
\begin{barticle}[author]
\bauthor{\bsnm{Hu},~\bfnm{Guanyu}\binits{G.}},
  \bauthor{\bsnm{Yang},~\bfnm{Hou-Cheng}\binits{H.-C.}} \AND
  \bauthor{\bsnm{Xue},~\bfnm{Yishu}\binits{Y.}}
(\byear{2021}).
\btitle{Bayesian group learning for shot selection of professional basketball
  players}.
\bjournal{Stat}
\bvolume{10}
\bpages{e324}.
\end{barticle}
\endbibitem

\bibitem[\protect\citeauthoryear{Jiao, Hu and Yan}{2021}]{jiao_21}
\begin{barticle}[author]
\bauthor{\bsnm{Jiao},~\bfnm{Jieying}\binits{J.}},
  \bauthor{\bsnm{Hu},~\bfnm{Guanyu}\binits{G.}} \AND
  \bauthor{\bsnm{Yan},~\bfnm{Jun}\binits{J.}}
(\byear{2021}).
\btitle{A {Bayesian} marked spatial point processes model for basketball shot
  chart}.
\bjournal{Journal of Quantitative Analysis in Sports}
\bvolume{17}
\bpages{77--90}.
\bdoi{10.1515/jqas-2019-0106}
\end{barticle}
\endbibitem

\bibitem[\protect\citeauthoryear{Malsiner-Walli, Fr{\"u}hwirth-Schnatter and
  Gr{\"u}n}{2016}]{malsiner2016model}
\begin{barticle}[author]
\bauthor{\bsnm{Malsiner-Walli},~\bfnm{Gertraud}\binits{G.}},
  \bauthor{\bsnm{Fr{\"u}hwirth-Schnatter},~\bfnm{Sylvia}\binits{S.}} \AND
  \bauthor{\bsnm{Gr{\"u}n},~\bfnm{Bettina}\binits{B.}}
(\byear{2016}).
\btitle{Model-based clustering based on sparse finite Gaussian mixtures}.
\bjournal{Statistics and computing}
\bvolume{26}
\bpages{303--324}.
\end{barticle}
\endbibitem

\bibitem[\protect\citeauthoryear{McLachlan}{2000}]{mclachlan2000finite}
\begin{barticle}[author]
\bauthor{\bsnm{McLachlan},~\bfnm{Geoffrey}\binits{G.}}
(\byear{2000}).
\btitle{Finite mixture models}.
\bjournal{A wiley-interscience publication}.
\end{barticle}
\endbibitem

\bibitem[\protect\citeauthoryear{{McMahan}}{2018}]{mcmahan}
\begin{bmisc}[author]
\bauthor{\bsnm{{McMahan}},~\bfnm{Ian}\binits{I.}}
(\byear{2018}).
\btitle{How (and why) position-less lineups have taken over the NBA playoffs}.
\bnote{\url{https://www.theguardian.com/sport/blog/2018/may/01/how-and-why-position-less-lineups-have-taken-over-the-nba-playoffs}}.
\end{bmisc}
\endbibitem

\bibitem[\protect\citeauthoryear{Muniz and Flamand}{2022}]{MunizFlamand}
\begin{barticle}[author]
\bauthor{\bsnm{Muniz},~\bfnm{Megan}\binits{M.}} \AND
  \bauthor{\bsnm{Flamand},~\bfnm{Tulay}\binits{T.}}
(\byear{2022}).
\btitle{A weighted network clustering approach in the NBA}.
\bjournal{Journal of Sports Analytics}
\bvolume{8}
\bpages{251--275}.
\end{barticle}
\endbibitem

\bibitem[\protect\citeauthoryear{Myers}{2020}]{BPM}
\begin{bmisc}[author]
\bauthor{\bsnm{Myers},~\bfnm{Daniel}\binits{D.}}
(\byear{2020}).
\btitle{About Box Plus/Minus (BPM)}.
\bnote{\url{https://www.basketball-reference.com/about/bpm2.html}}.
\end{bmisc}
\endbibitem

\bibitem[\protect\citeauthoryear{Nandram}{1998}]{nandram1998bayesian}
\begin{barticle}[author]
\bauthor{\bsnm{Nandram},~\bfnm{Balgobin}\binits{B.}}
(\byear{1998}).
\btitle{A Bayesian analysis of the three-stage hierarchical multinomial model}.
\bjournal{Journal of Statistical Computation and Simulation}
\bvolume{61}
\bpages{97--126}.
\end{barticle}
\endbibitem

\bibitem[\protect\citeauthoryear{Reich et~al.}{2006}]{reich2006spatial}
\begin{barticle}[author]
\bauthor{\bsnm{Reich},~\bfnm{Brian~J}\binits{B.~J.}},
  \bauthor{\bsnm{Hodges},~\bfnm{James~S}\binits{J.~S.}},
  \bauthor{\bsnm{Carlin},~\bfnm{Bradley~P}\binits{B.~P.}} \AND
  \bauthor{\bsnm{Reich},~\bfnm{Adam~M}\binits{A.~M.}}
(\byear{2006}).
\btitle{A spatial analysis of basketball shot chart data}.
\bjournal{The American Statistician}
\bvolume{60}
\bpages{3--12}.
\end{barticle}
\endbibitem

\bibitem[\protect\citeauthoryear{Sandholtz and Bornn}{2020}]{sand_born}
\begin{barticle}[author]
\bauthor{\bsnm{Sandholtz},~\bfnm{Nathan}\binits{N.}} \AND
  \bauthor{\bsnm{Bornn},~\bfnm{Luke}\binits{L.}}
(\byear{2020}).
\btitle{Markov decision processes with dynamic transition probabilities: {An}
  analysis of shooting strategies in basketball}.
\bjournal{The Annals of Applied Statistics}
\bvolume{14}.
\bdoi{10.1214/20-AOAS1348}
\end{barticle}
\endbibitem

\bibitem[\protect\citeauthoryear{Sandholtz, Mortensen and
  Bornn}{2020}]{sand_etal_20}
\begin{barticle}[author]
\bauthor{\bsnm{Sandholtz},~\bfnm{Nathan}\binits{N.}},
  \bauthor{\bsnm{Mortensen},~\bfnm{Jacob}\binits{J.}} \AND
  \bauthor{\bsnm{Bornn},~\bfnm{Luke}\binits{L.}}
(\byear{2020}).
\btitle{Measuring spatial allocative efficiency in basketball}.
\bjournal{Journal of Quantitative Analysis in Sports}
\bvolume{16}
\bpages{271--289}.
\bdoi{10.1515/jqas-2019-0126}
\end{barticle}
\endbibitem

\bibitem[\protect\citeauthoryear{{R Core Team}}{2022}]{r-man}
\begin{bmanual}[author]
\bauthor{\bsnm{{R Core Team}}}
(\byear{2022}).
\btitle{R: A Language and Environment for Statistical Computing}
\bpublisher{R Foundation for Statistical Computing},
\baddress{Vienna, Austria}.
\end{bmanual}
\endbibitem

\bibitem[\protect\citeauthoryear{Thomas and Ventura}{2015}]{hockey-vent}
\begin{bmisc}[author]
\bauthor{\bsnm{Thomas},~\bfnm{A.~C}\binits{A.~C.}} \AND
  \bauthor{\bsnm{Ventura},~\bfnm{S.~L.}\binits{S.~L.}}
(\byear{2015}).
\btitle{War on Ice}.
\end{bmisc}
\endbibitem

\bibitem[\protect\citeauthoryear{Williams et~al.}{2025}]{elmoresup}
\begin{barticle}[author]
\bauthor{\bsnm{Williams},~\bfnm{Benjamin}\binits{B.}},
  \bauthor{\bsnm{Schliep},~\bfnm{Erin~M}\binits{E.~M.}},
  \bauthor{\bsnm{Fosdick},~\bfnm{Bailey~K}\binits{B.~K.}} \AND
  \bauthor{\bsnm{Elmore},~\bfnm{Ryan}\binits{R.}}
(\byear{2025}).
\btitle{Supplement to "Expected Points Above Average: A Novel NBA Player Metric
  Based on Bayesian Hierarchical Modeling"}.
\end{barticle}
\endbibitem

\bibitem[\protect\citeauthoryear{Winston, Nestler and
  Pelechrinis}{2022}]{winston2022mathletics}
\begin{bbook}[author]
\bauthor{\bsnm{Winston},~\bfnm{Wayne~L}\binits{W.~L.}},
  \bauthor{\bsnm{Nestler},~\bfnm{Scott}\binits{S.}} \AND
  \bauthor{\bsnm{Pelechrinis},~\bfnm{Konstantinos}\binits{K.}}
(\byear{2022}).
\btitle{Mathletics: How gamblers, managers, and fans use mathematics in
  sports}.
\bpublisher{Princeton University Press}.
\end{bbook}
\endbibitem

\bibitem[\protect\citeauthoryear{Yin, Hu and Shen}{2022}]{yinhushen}
\begin{barticle}[author]
\bauthor{\bsnm{Yin},~\bfnm{Fan}\binits{F.}},
  \bauthor{\bsnm{Hu},~\bfnm{Guanyu}\binits{G.}} \AND
  \bauthor{\bsnm{Shen},~\bfnm{Weining}\binits{W.}}
(\byear{2022}).
\btitle{Analysis of Professional Basketball Field Goal Attempts via a Bayesian
  Matrix Clustering Approach}.
\bjournal{Journal of Computational and Graphical Statistics}
\bvolume{00}
\bpages{1--12}.
\end{barticle}
\endbibitem

\bibitem[\protect\citeauthoryear{Yurko, Ventura and Horowitz}{2018}]{nflwar}
\begin{barticle}[author]
\bauthor{\bsnm{Yurko},~\bfnm{Ronald}\binits{R.}},
  \bauthor{\bsnm{Ventura},~\bfnm{Samuel}\binits{S.}} \AND
  \bauthor{\bsnm{Horowitz},~\bfnm{Maksim}\binits{M.}}
(\byear{2018}).
\btitle{nflWAR: a reproducible method for offensive player evaluation in
  football}.
\bjournal{Journal of Quantitative Analysis in Sports}
\bvolume{15}
\bpages{163--183}.
\end{barticle}
\endbibitem

\bibitem[\protect\citeauthoryear{Zhang et~al.}{2018}]{zhang2018clustering}
\begin{barticle}[author]
\bauthor{\bsnm{Zhang},~\bfnm{Shaoliang}\binits{S.}},
  \bauthor{\bsnm{Lorenzo},~\bfnm{Alberto}\binits{A.}},
  \bauthor{\bsnm{G{\'o}mez},~\bfnm{Miguel-Angel}\binits{M.-A.}},
  \bauthor{\bsnm{Mateus},~\bfnm{Nuno}\binits{N.}},
  \bauthor{\bsnm{Gon{\c{c}}alves},~\bfnm{Bruno}\binits{B.}} \AND
  \bauthor{\bsnm{Sampaio},~\bfnm{Jaime}\binits{J.}}
(\byear{2018}).
\btitle{Clustering performances in the NBA according to players’
  anthropometric attributes and playing experience}.
\bjournal{Journal of sports sciences}
\bvolume{36}
\bpages{2511--2520}.
\end{barticle}
\endbibitem

\end{thebibliography}

\end{document}